# Compound Super-oscillation Lens for Reflective Confocal Imaging

**Pengcheng Zheng,**[1,3], **Zhaoxiang Zhu,**[2] **Xiangcan Pei,**[1,3] **Qinfei Wu,**[2] **Haowen Liang,**[2] **Yujie Chen,**[2] **Juntao Li,**[2] **and Xiangsheng Xie**[1,3,*]

[1] *Department of Physics, College of Science, Shantou University, Shantou, Guangdong 515063, China*

[2] *State Key Laboratory of Optoelectronic Materials and Technologies, School of Electronics and Information Technology, School of Physics, Sun Yat-sen University, Guangzhou 510275*

[3]*Key Laboratory of Intelligent Manufacturing Technology of MOE, Shantou University, Shantou, Guangdong 515063, China*

*\*Corresponding author: xxs@stu.edu.cn*

**Abstract:** The super-oscillation lens (SOL) can achieve super-resolution focusing but have to trade-off with weaker hotspots and higher sidebands. We propose a single compound SOL to achieve reflective confocal imaging in principle without additional lenses. The designed SOL consists of an outer lens and an inner lens which play the role of focusing lens and collective lens respectively. As a result, focusing and collecting functions can be simultaneously realized. The improved system can achieve excellent imaging performance with an ultra-high resolution (<0.34$\lambda$/NA, NA stands for numerical aperture), and almost negligible side lobe ratio and no side bands, which proved superior to conventional laser scanning confocal microscopy and single SOL. This technology can be attractive for a variety of applications in super-resolution imaging and biomedical sciences.

## 1. Introduction

Optical microscopy is a powerful detection tool in biotechnology and life science, but the resolution of a conventional optical imaging system is inherently limited by the diffraction limit. The development of super-resolution fluorescence microscopy (SRFM) provided effective solutions to some extent. For instance, stimulated emission depletion microscopy (STED), structured illumination microscopy (SIM), photoactivated localization microscopy (PALM), and stochastic optical reconstruction microscopy (STORM) can provide higher resolution of tens of nanometers in the far-field by distinguishing the fluorophores[1-4]. By introducing special fluorescent probes, SRFM can achieve a resolution of several tens of nanometers and has become a major tool. However, these SRFMs exhibit their powerful performance only when the target functional groups can be selectively labeled without photobleaching. Meanwhile, the process of sample preparation or labeling with multiple fluorescence simultaneously can be time-consuming and challenging.

A far-field label-free super-resolution imaging method using extremely rapid local variations in electromagnetic fields was recently established. The mathematical idea behind this is that a function can locally oscillate much faster than its fastest Fourier components [5] when precisely tailored. This imaging method has a novel name: super-oscillation imaging. Huang et al. provided the first demonstration of the optical super-oscillation phenomenon by diffraction from a quasiperiodic array of nanoholes [6]. Qiu [7] also considered that for a focusing lens with NA=nsinα, the size of the circularly symmetrical optical spot is determined by the highest spatial frequency kr=ksinα, where n is the refractive index of the medium after the lens, and α is the angle between the optical axis and the wavevector. The spot size corresponds to the distance between the central peak and the first zero of the zero-order Bessel function of the first kind ($J_0(2\pi r sin\alpha/\lambda)$), which yields a value of 0.38λ/NA and is close to the FWHM (full width at half maximum) in most cases. According to this analysis, 0.38λ/NA is the criterion for optical super-oscillation focusing. Kozawa et al. showed that a radially polarized Laguerre-Gaussian mode has the inherent ability to form super-oscillation spots simply by controlling the incident beam size[8]. The superoscillation lens[9], a two-dimensional electromagnetic metamaterial, consists of an array of subwavelength nanostructures, which can arbitrarily manipulate the phase, polarization, and amplitude of electromagnetic waves[10-13]. Compared with traditional optical devices, it has the advantages of ultra-thin thickness, compact structure, and easy integration. It has a wide range of applications in the field of holographic imaging[14-18], lens imaging[19-26], and polarization control[27-29], and attract great attention in recent years. However, a major drawback is the inevitable existence of large sidelobe intensity and high-energy sidebands close to the hotspot, which leads to trade-offs among the spot FWHM (full width at half maximum), FOV (field of view), and SNR (signal-to-noise ratio) of the optical super-oscillatory field, especially when the spot size is much smaller than the diffraction limit. If managed unsuitably, these sidelobes and sidebands of super-oscillation can drown the superoscillating signal [30].

These disadvantages have a bad effect on the practical application of super-oscillation. Previous research proposed various methods for compressing sidelobes and managing the outside bands of super-oscillation. For instance, the outside bands of super-oscillation can be moved away from the subwavelength hotspot owing to an appropriate window for imaging [31,32]. In another approach suggested that the outside bands of super-oscillation can remain with low energy for a weak subwavelength super-oscillation hotspot [33,34]. Several techniques were introduced using two photons to suppress the super-oscillation outside bands in confocal scanning fluorescence microscopy [35]. Recently, a hybrid between super-oscillation and subtraction was introduced to remove the outside lobes of the super-oscillation [36]. This approach requires two images for imaging processing, which makes the method more time-consuming. Indeed, dealing with super-oscillation sidelobes and sidebands remains a crucial consideration.

Confocal laser scanning microscopy (CLSM) is a popular technique that can obtain three-dimensional images of samples with high resolution [37]. This has the advantages of depth selectivity and noise filtering by recalculating the light field carrying the object information and pinhole filtering. Because of its strong compatibility, confocal microscopy can be considered a natural supplement to optical super-oscillation imaging, which can effectively filter out high-energy sidebands and suppress the sidelobe of the super-oscillation in the confocal plane. Rogers et al. used a scanning mode with super-oscillation lens (SOL) illumination, where the signal used to reconstruct the image was taken from the central part of the CCD camera, the region where the hotspot is projected in the absence of the object. This imaging tactic, also employed in confocal microscopy, allows for the removal of unwanted scattering from the halo[38]. However, a single SOL cannot collect the scattering light, and the objective lens must be introduced for imaging. Therefore, there are few reports on improving the imaging quality of SOL using confocal technology [39-41].

On the basis of these insights, we demonstrate a technique to improve the CLSM by using a compound SOL (CSOL) consisting of an outer lens and an inner lens, acting as a focusing lens and a collective lens, respectively. The improved system can achieve ultra-high resolution of 0.315λ/NA in the far field of 1500λ, with almost negligible sidelobe ratio (< 2%) and without side-bands. The imaging performance of CSOL with large NA (0.9285) is similar. To the best of our knowledge, it is the first time to realize reflective CLSM using a single CSOL without high NA objective lens. We believe that the compound design of SOL will open a new way to effectively solve the contradiction between resolution, signal-to-noise ratio and field of view of conventional SOL.

## 2. Theory and design

Optical super-oscillation refers to the phenomenon in which a band-limited function contains local oscillations faster than the fastest Fourier components [42]. For optical waves, the optical super-oscillation phenomenon is the delicate interference of the far-field propagating waves, and it can be engineered to achieve a sub-diffraction-limit-focusing hotspot without the contribution of evanescent waves [43,44]. Among the optical super-oscillatory features, phase distribution plays a crucial role [45,46]. The optical field can be described by $E(r) = A(r)\,exp[i\varphi(r)]$, where $A(r)$ and $\varphi(r)$ denote the amplitude and phase, respectively. By substituting E(r) into the Helmholtz equation, we obtain

$$\begin{cases} \nabla^2 \varphi(r) + \nabla[lnA^2(r)] \cdot \nabla\varphi(r) = 0 \\ \nabla^2 \varphi(r) + [k^2 - |\nabla\varphi(r)|^2]A(r) = 0 \end{cases} \quad (1)$$

where $k$ is the wavenumber, and $\nabla\varphi(r)$ is the local wavenumber expressed as the gradient of the phase distribution. When the length of $\nabla\varphi(r)$ exceeds the wavenumber k at the point $\vec{r}$, there is a fast decay in the neighboring area, which leads to the formation of superoscillatory fields. Theoretically, super-oscillation allows the formation of arbitrarily small optical features, which can be used for super-resolution focusing and imaging, but this is followed by a smaller FOV and a larger sidelobe ratio.

Confocal imaging techniques such as confocal fluorescence microscopy or confocal 2-photon microscopy are based on point-by-point raster scanning imaging. As shown in Fig. 1(A), a point-like light source is focused on a lens or an objective onto a sample. The spatial extension of the focus spot on the sample is determined by the wavelength λ and the quality of the image formation. The image spot is refocused through the same (or a second) lens onto an aperture (pinhole) in front of the detector. The size of the pinhole is chosen such that only the central part of the focus can pass through the pinhole and reach the detector. Rays that do not come from the focal plane are unable to pass through the pinhole. Therefore, the image contrast is strongly enhanced.

One of the most important properties of a confocal system is the point spread function (PSF), which is mainly determined by the microscope objective. The resolution of CLSM can, therefore, be modifying the spatial arrangement of the excitation and detection PSFs [47], which is called "PSF engineering". We have proposed fine light field modulation techniques to enhance the optical resolution of a CLSM by harnessing both excitation and detection PSFs. The lateral resolution of 1/5λ (less than the super-oscillation criterion) is obtained in the far-field with a working distance greater than ∼500λ [48].

The image formation of a confocal microscope can be described with a coherent light source as [49]

$$\vec{E}_{imag}(\rho,\varphi,z) = \frac{\omega^2}{\varepsilon_0 c^2} \overleftrightarrow{G}_{PSF}(\rho,\varphi,z) \cdot \vec{\alpha} \vec{E}_{exc} \qquad (2)$$

where $\overleftrightarrow{G}$ is the dyadic point-spread function, $\vec{E}_{exc}$ is the light field distribution of the excitation field, and α is the polarizability tensor [50]. When the interaction between $\vec{E}_{exc}$ and the sample is linear, Eq.(2) can be simplified as $I(x_i, y_i) = |\iint h_1(x_0, y_0) h_2(x_0, y_0) t(x_i - x_0, y_i - y_0) \, dx_0 dy_0|^2$, where $h_1$ represents the excitation PSF, and $h_2$ represents the detection PSF. For simplification, both functions should be scalar values. The product $h_{eff} = h_1 h_2$ is called the effective PSF of the optical system. By choosing an appropriate pinhole diameter, the lateral resolution can be increased by up to a factor of $\sqrt{2}$.

In combination with the optical super-oscillation theory and CLSM, we propose a compound lens to improve the conventional CLSM [Fig. 1(B)]. The compound lens consists of an outer lens and an inner lens, which replace the objective lens (corresponding to the excitation PSF) and the cluster lens (corresponding to the detection PSF) in the CLSM, respectively [Fig. 1(C)]. The radius of the compound lens was 100 μm and the duty ratio was 85%, thus the radius of the inner lens was 170 μm. In addition, both inner and outer lenses were both designed as binary phase-mask-based SOLs working at 632.8nm wavelength region [50]. In terms of performance, the outer lens is used to produce as small a light focal spot as possible, and inner lens is used to suppress the side lobe level. The combination of the two can comprehensively improve the imaging quality of the system, such as resolution, signal-to-noise ratio and field of view. The structure of the lens is shown in Fig. 1(D). The upper part of Fig. 1(D) is a top view of the lens, which consists of a series of concentric rings. The width of each ring was a fixed value of T. Owing to symmetry, only the upper half of the lens is illustrated in the figure.

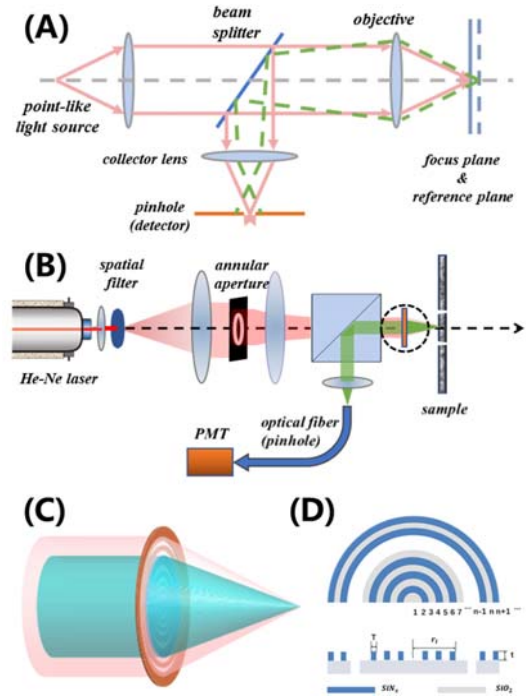

Fig. 1. Schematic diagrams of (A) traditional CLSM, (B) the designed microscopy setup (Red represents incident light, and green represents reflected light), (C) focusing and collection of the CSOL and (D) Geometrical structure of binary amplitude-phase lens with top view (upper) and cross-section of the lens ((lower).

As shown in Fig. 1(D), the rings comprise two types of layers: a glass ($SiO_2$) layer and a $SiN_x$ layer grown on a glass substrate. The former was used to build the underlayer to facilitate processing, and the latter was used to achieve binary phase modulation. This structure was chosen for two reasons: first, silicon nitride has a higher transmittance than plasma metamaterials; secondly, the cylindrosymmetric rings structure is easier to fabricate. The width and thickness of the rings were T and t, respectively. The binary phase was realized by controlling the $SiN_x$ stripe thickness 0 and t for phase changes of 0 and $\pi$, respectively. The value of t was obtained by $t = \lambda/2(n_{SiN_x} - 1)$, where $n_{SiN_x}$ is the refractive index of $SiN_x$. We set T = 200 nm and t = 323 nm.

### 3. Result and discussion

### 3.1 Design and fabrication of CSOL

Super-oscillation is the result of the interference of coherent light. Therefore, accurate calculation of the diffraction light field distribution is key to constructing a superoscillating light field and the design of superoscillating optical devices, the commonly used method is vector angular spectrum theory(VAS) [51,52]. Compared with the finite-difference time-domain method and the finite element method, VAS can be completed by a fast Hankel transformation [53-55], which has a faster computing speed and is more conducive to designing SOLs with large areas and complex structures.

The design of an SOL mainly relies on optimization algorithms, among which particle swarm algorithms [56,57] are the most commonly used. With the particle swarm optimization approach, the SOL design process can be simple and easily understood (full details are provided in Appendix A).

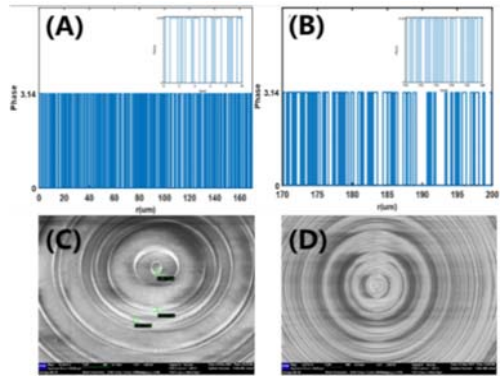

Fig. 2. Phase distribution of inner SOL(A) and outer SOL(B). SEM images of the CSOL (C), (D).

The optimized phase distribution is shown in Fig.2(A), (B). To describe the CSOL (containing 1000 rings) more concisely, the transmission is coded from the first ring (innermost) to the n-th ring (outermost) by continuously converting every four successive binary digits into one hexadecimal digit (Table 1). For example, the transmissions of the first four rings, "1000," were coded to be "8", the second hexadecimal digit "F" denotes the real transmissions of "1111." The digit "0" for $SiO_2$ and the digit "1" for $SiN_x$. The phase distribution of the inner and outer SOLs after optimization, containing details and parts, is shown in Fig. 2.

To verify the design, a circular $SiN_x$ SOL with diameter of 400um was fabricated on a quartz substrate with 323 nm $SiN_x$. The pattern is defined in a positive resist (ARP 6200) by electron beam lithography (EBL, Raith Vistec EBPG-5000plusES) written at 100 keV (full details are provided in Appendix B). The patterned sample is etched in a Reactive-ion etching (RIE, Oxford Instruments Plasma Pro 100RIE), which has been optimized for vertical, smooth sidewalls etching of $SiN_x$. Scanning electron microscope (SEM, Zeiss Auriga) images of the CSOL are shown in Fig. 2(C), (D).

**Table 1. Transmittance functions of the optimized SOLs**

| Number of ring slits | Optimization result of the transmittance value |
|---|---|
| #1 - #200 | 8F93DE217851C9336076FC7474CDE991D05E4768BD785CFAE8 |
| #201 - #400 | 2BCF443B45133D9C5E85CA99151BF1E9405715EA7C11DDA24E |
| #401 - #600 | B36C989A7DCC45B50F3E6A726AF6BF197F95C0CE37F4135F22 |
| #601 - #800 | E9065FDC5127812373DB8F6E2A42DF6582B22067D51E3285D9 |
| #801 - #1000 | BC23E8CAFA9C65AEB55962D598D15C3EB0D98052022538DC41 |

## 3.2 Simulation and characterization of COSL

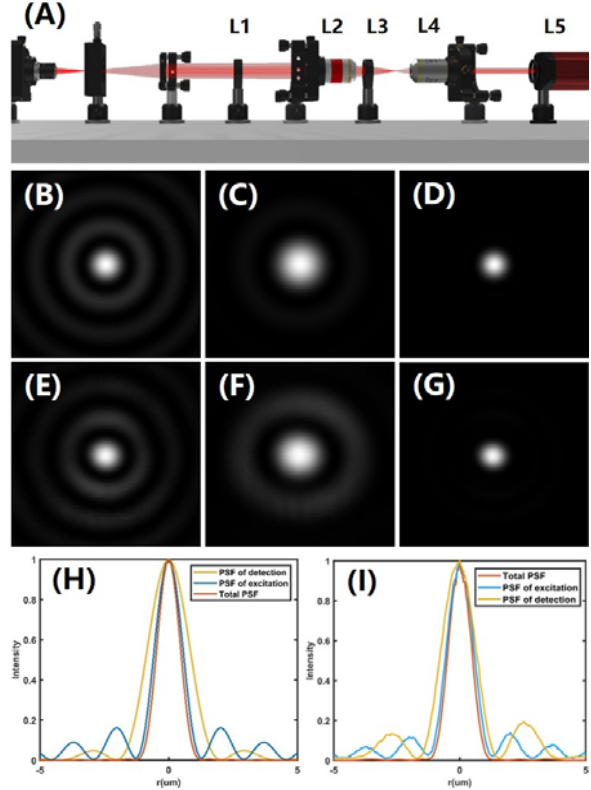

Fig. 3. (A) Measurement set-up used to characterize the CSOL according to the design shown in Table 1. (L1): annular aperture (inradius=850μm, exradius = 1000μm); (L2): 5×objective (NA=0.13); (L3): CSOL ; (L4): 100×objective (NA=0.6); (L5): CMOS camera at 632.8nm wavelength. (B)~(D) Simulated excitation PSF(B, outer SOL), detection PSF(C, inner SOL) and total PSF(D). (E)~(G) Experimentally measured excitation PSF (E, outer SOL), detection PSF (F, inner SOL) and total PSF(G). (H) Simulated intensity distribution of excitation PSF (blue), detection PSF (yellow) and total PSF (red). (I) Experimental intensity distribution of excitation PSF (blue), detection PSF (yellow) and total PSF (red).

The focusing performance of outer SOL and inner SOL is evaluated based on numerical simulation [(Fig. 3 (B)-(C)]. According to the operating principle of CLSM, we calculate the focal spot intensity profile of the CSOL [Fig. 3 (D)]. Table 2 shows the FWHM and SLR (sidelobe ratio) of the excitation, detection, and total PSFs of this system. The outer lens achieves far-field the super-diffraction focusing with a FWHM of 0.372λ/NA, which is derived from the super-oscillation effect of local spatial frequency, but there is a high sidelobe level (43.39%). The SLR of inner SOL is as low as 8.79%, but the resolution was not good enough. CSOL has both the advantages of the previous two, FWHM and SLR are 0.315λ/NA and 0.89% respectively. By comparing the excitation, detection, and total PSFs of the CLSM in Fig. 3(H), it can be found that the maximum of the exciting PSF is at the minimum of the detection PSF.

This results in the system's sidelobe being suppressed well. We believe this phenomenon is meaningful because for an SOL, high resolution is often accompanied by a large sidelobe, but when it is combined with a confocal system, it can improve the system resolution without being limited by a large sidelobe ratio. The diffraction limit is based on the Airy disk, which is the intensity profile generated by a circular aperture in the focal plane of an objective lens and has a FWHM=0.51$\lambda$/N.A. When the circular aperture is changed by an annular aperture, it is obtained another intensity profile that has a FWHM below the Airy disk [58]. The maximum reduction of this FWHM is achieved when the radius of the annular aperture approaches the radius of the circular aperture (similar to the super-oscillatory criterion) and it is, approximately, 30%. In our case as the quotient between the radius is 85%, the reduction of the FWHM of the intensity profile is approximately 20%(FWHM=0.489$\lambda$/NA). We used ONSOL [33,34] to compress FWHM to 0.372$\lambda$/NA on the excitation point spread function (Table 3).

The characterization of the focusing performance is shown in Fig.3 (A), we use an aperture L1 and a low-power objective lens L2 to produce an illumination beam matching CSOL's Inner SOL and Outer SOL. The focal points of inner SOL and outer SOL are dPSF and ePSF. The total point spread function is their product. Since the focal spot size is only about 1um, and the minimum resolution of CMOS is 5.04um, the focal spot was enlarged by 100 times with a 100x objective lens and detected by the CMOS [Fig. 3(E)~(G)]. FWHM and SLR measured experimentally are shown in Table 2. The FWHM and SLR of the total PSF are 0.314$\lambda$/NA and 1.87%, respectively. And the relationship between the excitation PSF, detection PSF, and total PSF is almost consistent with the numerical simulation [Fig. 3(I)]. In terms of total PSF, the difference in FWHM and SLR between experimental value and calculated value are 3.23nm (0.001$\lambda$/NA) and 0.98%. This subtle discrepancy can be explained by fabrication tolerances and measurement error.

An interesting phenomenon in the experiment caught our attention. The focal spot is still recognizable when the illumination ring is displaced 3μm off the axis, though the sidelobes become asymmetric. And there is very little change in FWHM. Hence slight distortions (slight annular aperture displacements) are tolerable. This design can effectively improve the fault-tolerant ability in the measurement process.

In the direction of light transmission, the numerical simulation results show that inner SOL has a significant compression effect on the focal depth of the system, and the final focal length is 1000μm, which is completely consistent with the preset value (full details are provided in Appendix C).

Therefore, we demonstrate the feasibility of combining CSOL and CLSM to achieve super-oscillatory focusing with a FWHM of 0.315 $\lambda$ /NA and a SLR < 2%, and no sidebands.

**Table 2. FWHM and SLR of excitation PSF (outer SOL), detection PSF (inner SOL) and total PSF**

|  |  | excitation PSF | detection PSF | total PSF |
|---|---|---|---|---|
| Simulation | FWHM | 0.372 λ /NA | 0.476λ /NA | 0.315 λ /NA |
|  | SLR | 43.39% | 8.79% | 0.89% |
| Experiment | FWHM | 0.372 λ /NA | 0.475 λ /NA | 0.314 λ /NA |
|  | SLR | 38.84% | 17.42% | 1.87% |

Table 3 Simulation of comparison of focusing performance between ONSOL and annular aperture

|  | FWHM | SLR |
|---|---|---|
| ONSOL | 0.372λ/NA | 43.39% |
| annular aperture | 0.489λ/NA | 37.83% |

### 3.3 Simulation and experiment of imaging

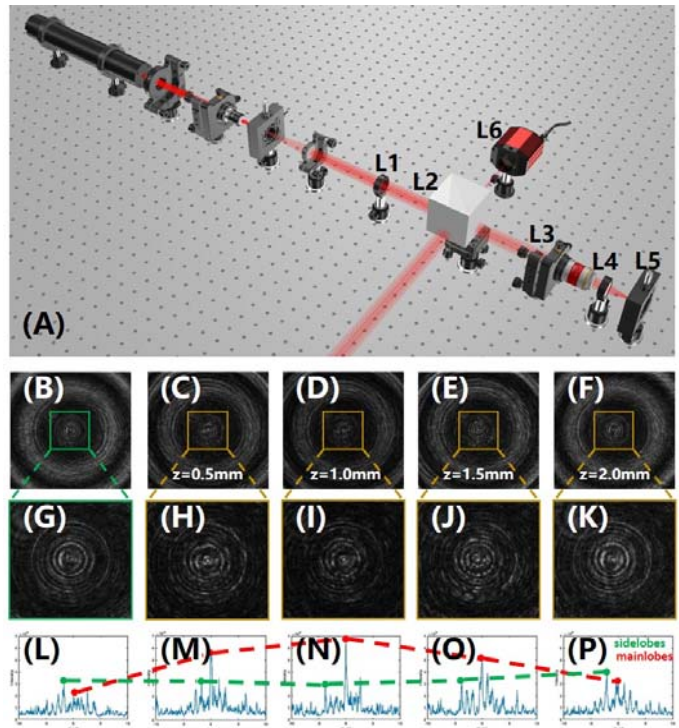

Fig. 4. (A) Diagram of reflector confocal system for scanning imaging. (L1): annular aperture (inradius=850um, exradius = 1000um); (L2): Beam splitting prism; (L3): 5×objective (NA=0.13); (L4): CSOL; (L5): piezo-stage; (L6): photomultiplier. Intensity profile without sample[(B), (G)] and when the plane mirror is placed 0.5mm [(C), (H)], 1mm [(D), (I)], 1.5mm [(E), (J)] and 2mm [(F), (K)] away from CSOL. (L)~(P), Corresponding intensity distribution.

In order to illustrate the imaging capability of CSOL, a reflective confocal system as shown in Fig. 4(A) was set up and this system was used to scan and image the target sample (resolution test target). The detailed imaging process is as follows: After passing through the annular aperture (inradius=850um, exradius = 1000um), the parallel beam is focused on the CSOL by the objective lens, forming the annular illumination (inradius=170um, exradius = 200um). The illumination ring is located in the region of outer SOL of CSOL. The focal spot of the outer SOL is irradiated on the sample (reflecting resolution test target), and the reflected signal is collected by the inner SOL of CSOL, finally detected by PMT (photomultiplier) after passing through the objective lens and splitting prism. The sample (reflection resolution test target) is fixed on the piezo-stage. Therefore, we can control the movement of piezo-stage to perform point-by-point scanning of the sample (the reflective resolution test target).

At first, we had an unavoidable problem, which was that the system was almost impossible to image because of the high background noise. When we removed the target sample, a relatively strong optical signal was still detected due to light reflected from the CSOL substrate [Fig. 4(B), (G)]. Next, we placed the plane mirror at 0.5mm, 1mm, 1.5mm and 2mm away from the sample successively, and captured the corresponding image of the sample [Fig. 4(C)~(F)]. A comparison of Fig. 4(L)~(P) shows that the intensity of side lobe (green dotted line) is independent of the object distance, but the strength of the main lobe (red dotted line) varies with the object distance and reaches its maximum value when the plane mirror is in the focal plane of CSOL. These results suggest that the background noise is due to the reflection of the annular illumination caused by the air/ glass interface.

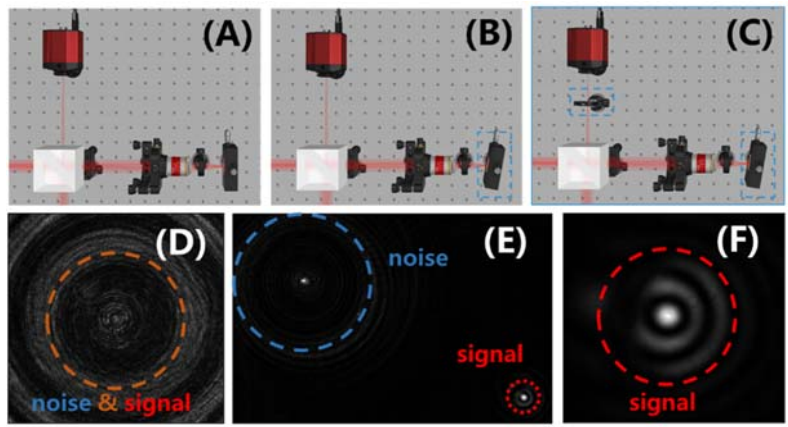

Fig. 5. The original imaging device(A) and imaging results(D); Device after sample tilting (B) and imaging results(E); The device(C) and imaging(F) results after tilting sample and adding a circular aperture.

Next, we show the noise filtering method in Fig. 5. When we use the device shown in Fig. 4(A) to image a reflective object (a plane mirror) [Fig. 5(A)], there is strong back-reflected noise, which is the result of sharp edge diffraction of light reflected from the CSOL (compound super-oscillation lens) substrate, and it overwhelms the signal carrying information about the object [Fig. 5(D)]. To remove noise, we first tilt the sample (the resolution test target) at an appropriate Angle [Fig. 5(B)], which separates the detection signal from the back-reflected noise [Fig. 5(E)]. Then, a circular aperture is used to block out most of the back-reflected noise [Fig. 5(C)], and a relatively complete detection signal is finally extracted [Fig. 5(F)]. Obviously, this is a compromised experimental scheme, which can only filter out most but not all of the back-reflected noise. In addition, the tilted sample will also lead to energy imbalance of the detection points and decreased resolution. The actual deflection Angle we set is about 5.7°. In this case, the detection points are clearly visible and the resolution decrease is not significant.

Fig. 6(E)~(G) is the scanning imaging result of the resolution test target using the improved system. We tested the resolution of this system for an angle of the rectangle in Fig. 6(I), and the result is 1.12μm, which is only 0.1um different from the simulated result of 1.02μm (0.315λ/NA). This difference can be interpreted as the off-axis deviation of target sample, the interference of residual background noise and measurement error. The imaging results of CLSM are shown in Fig. 6(B)~(D). By comparison, the CLSM has a resolution of 1.82μm [Fig. 6(H)], which is significantly worse than CSOL(1.12μm). The above results show that CSOL combined with confocal system has been proved experimentally to achieve a resolution of 1.12μm at far field of 1500λ, which is close to the simulation result of 1.02μm (0.315λ/NA) and obviously better than the traditional CLSM(1.82μm).

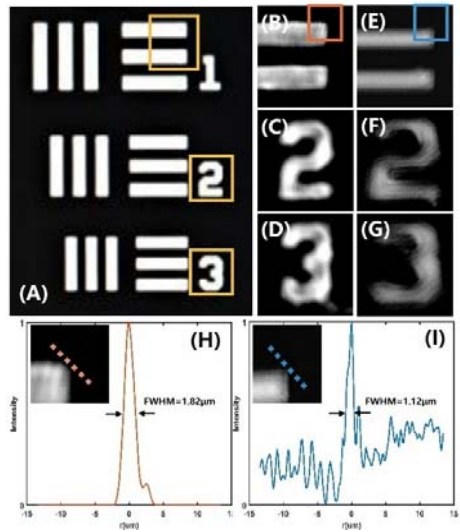

Fig.6. (A) Schematic diagram of samples (resolution test target). (B)~(D) Conventional CLSM imaging results of samples (resolution test target). (E)~(G) CSOL imaging results of samples (resolution test target). (H)~ (I) Typical translationally resolved measurement results of CLSM(H) and CSOL(I).

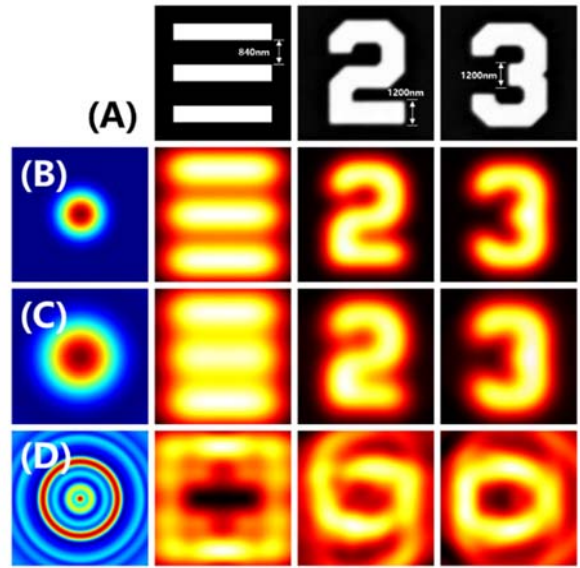

Fig. 7. Numerical comparison of imaging results with different lens system. (A) target objects (lines, "2"-shaped and "3"-shaped). (B) imaging by the proposed CSOL. (C) imaging by conventional CLSM and (D) imaging by single super-oscillation lens. The first column shows their effective PSFs

Fig. 7 shows further numerical simulation demonstration. We simulated the imaging effect of this system on different samples, such as lines objects, "2"-shaped objects and "3"-shaped objects [Fig. 7]. The images expected for different objects were simulated by a two-dimensional convolution of the PSF [Fig. 7(B)] and object [Fig. 7(A)]. For comparison, the imaging results of the conventional CLSM [Fig. 7(C)] and a single SOL [Fig. 7(D)] were also simulated. Concentrating on the imaging results of the lines objects, the CSOL has the best imaging results, and rectangles are insufficient to be distinguished clearly by conventional CLSM. For a single super-oscillating lens, high-energy side-bands of super-oscillation can drown out the super-oscillating signal, resulting in poor imaging results. In addition, we compare the imaging results of "2"-shaped object and "3"-shaped object. It is not difficult to see that the result of CSOL is clearest. The "2"-shaped object and "3"-shaped object can also be distinguished by conventional CLSM, but there is stronger edge diffraction leading to less clear edges, as spatial resolution of conventional CLSM is not high enough. When a single SOL is used, the object cannot be observed at all, due to the strong noise caused by the high-energy sidebands. In general, CSOL is better at resolving object details than CLSM, and effectively solve the contradiction between FWHM and SLR of conventional SOL.

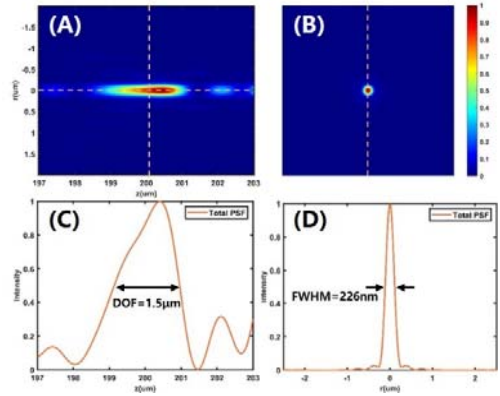

Fig. 8. Simulated focal spot intensity distribution of the large NA CSOL: axial intensity profile (A) and distribution (C), transverse intensity profile (B) and distribution (D).

For the case of large numerical aperture, we also designed a CSOL with a radius of 500um and a focal length of 200um, corresponding to a numerical aperture NA=0.9285. The difference is that this CSOL is based on radially polarized beam incident, because Relevant studies show that the focal spot corresponding to the large numerical aperture appear to be elliptic ones when linearly polarized beam incident [47]. Numerical simulation results show that the CSOL can achieve a focusing spot with a FWHM of 226nm (0.33λ/NA), a SLC of 6.04%, and a focal depth of 1.5μm (as shown in Fig. 8.). This indicates that CSOL can also achieve almost sidelobe-free super-oscillation focusing in the near field.

Because the focal depth is short in the case of large numerical aperture, it is difficult to filter the back-reflected noise by tilting the sample. Moreover, the step size of the current piezo-stage is tens of nanometers, which is not enough to provide accurate scanning accuracy. Therefore, we use numerical simulation method to characterize the imaging performance of the system. The simulation results of imaging effect are shown in Fig. 9. Similar to the far-field situation shown in Fig. 7, the imaging quality of CSOL in the near-field is higher than that of conventional CLSM and single SOL. The resolution of conventional CLSM is insufficient, while single SOL has low signal-to-noise ratio and small field of view. As shown in Fig. 9 (D), the effective signal of a single SOL is easily drown by high-energy sidebands, so that only noisy signals can be observed.

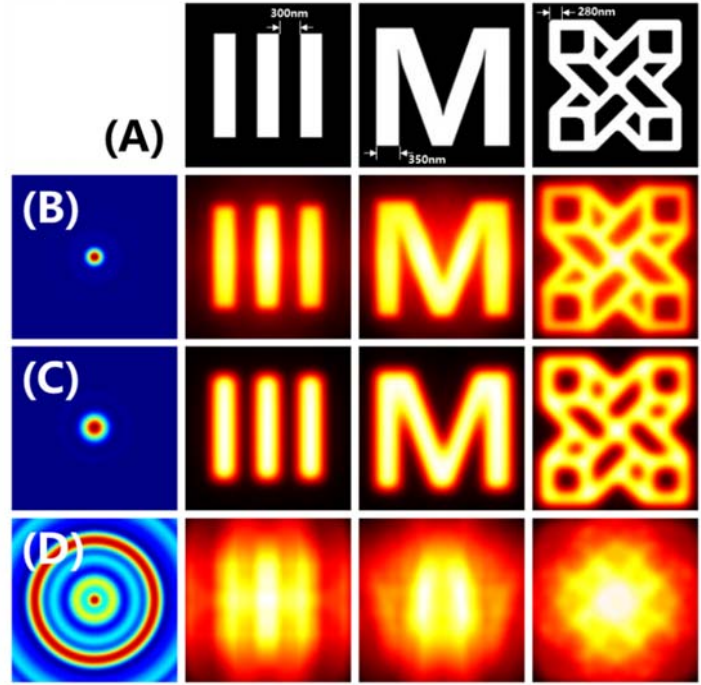

Fig. 9. Numerical comparison of imaging results with different lens system in the case of high NA. (A) target objects. (B) imaging by the proposed CSOL. (C) imaging by conventional CLSM and (D) imaging by single super-oscillation lens. The first column shows their effective PSFs

## 4. Conclusion and prospect

To summarize, we presented the first demonstration of a CSOL based on SOL in detail for reflective confocal system, which achieves a truly super-oscillation imaging without additional collecting lens. It consists of an inner SOL and an outer SOL. Outer SOL produces an extremely small spot whose width is below the super-oscillation criterion. Inner SOL is used to collect reflected light that carries information about the sample, further compress the focus spot and weakening the side-lobes.

The highlights of this work are as follows: Firstly, we optimize the resolution, signal-to-noise ratio and field of view of the SOL comprehensively by combining with the reflective confocal system for the first time. The imaging performance of the conventional SOL needs to be weighed among the three factors [45]. Secondly, we realize the reflection confocal microscopy experimentally based on SOL for the first time. Conventional transmission scanning schemes combined with SOL need to collect signals by an objective lens with a high numerical aperture, but our scheme only needs a single SOL to achieve the functions of excitation and detection [40]. The system has significant advantages over transmission CLSM in terms of volume, integration, and flexibility.  Finally, for the first time, we use a SOL to image a reflective sample rather than a hollow one [59] or fluorescent nanoparticles [60]. Be that as it may, we believe there is still room for improvement in our work. For instance, a radially polarized beam, which is more compact and focused, can be used as the light source, and V-shaped metamolecules [10] rather than rings can be used as superoscillating elements to carefully design quasicontinuous phase modulation. The scheme can also be extended to confocal fluorescent microscopy. We believe that the present method has advantages in versatility, as it provides a new route for improving the resolving power of confocal systems and overcoming the limitations of the significant nearby sidelobes and sidebands of the super-oscillating lens. This technology can be attractive for a variety of applications in super-resolution imaging and biomedical sciences.

**Appendix A: Structural design of CSOL**

The design procedure is described in the flowchart in Fig. 10. First, for specified parameters of the target field such as the FWHM and sidelobe ratio, a group of lenses with different transmission functions is randomly generated. Then, the diffraction pattern of each lens is calculated on the target focal plane for the specified incident waves. A fitness function is calculated for each lens by comparing the diffraction field and target fields. Finally, the transmission function of each lens is updated according to the fitness function. This procedure is repeated until the best fitness function attains a predefined value.

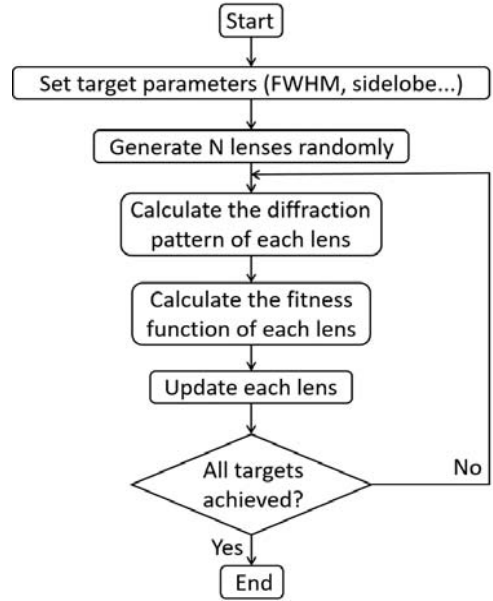

Fig. 10. Flowchart of particle swarm algorithm for SOL optimization.

In this work, we set the fitness function to

$$fitness = m \cdot \frac{FWHM - targetFWHM}{targetFWHM} + n \cdot \frac{SLC - targetSLC}{targetSLC} \qquad (3)$$

where, FWHM and SLC respectively represent the actual focal spot width and sidelobe ratio, while targetFWHM and targetSLC are their corresponding target values, m and n are the weight factors.

For outer SOL, to provide high scan resolution, we set $targetFWHM = \frac{0.38\lambda}{NA}, targetSLC = 50\%$, and m>n. For inner SOL, to suppress the background noise effectively, we set $targetFWHM = \frac{0.51\lambda}{NA}, targetSLC = 10\%$, and m<n. Finally, when the fitness value is less than zero, we can obtain the qualified transmittance distribution.

**Appendix B: Electron beam lithography process**

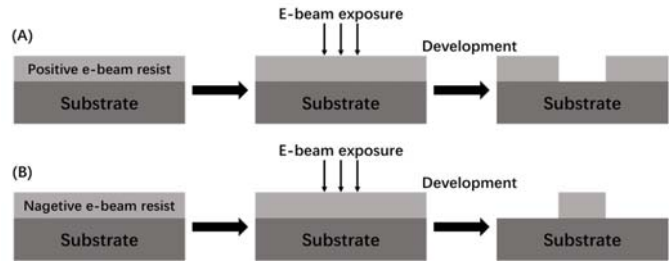

Fig. 11. Flow chart of electron beam lithography process. (A) Positron e-beam resist; (B) Negative e-beam resist

EBL (Electron beam lithography) can directly draw the map pattern on the substrate surface by using the electron gun of SEM (Scanning electron microscope), so it is a photolithography method without mask plate, and does not need to go through the alignment process of mask plate. EBL has higher precision than ordinary lithography, even down to 10 nm. As shown in Fig. 11, the EBL process similar to ultraviolet light moment: Firstly, resist (positron resist or negative resist) is applied to the substrate and then impinged with electrons fired from an electron gun in an electron beam lithography machine. In this case, the resist will affect the solubility of the material in the developer. When positive resist (PMMA, ARP6200, etc.) is used, the solubility increases (Fig. 11(A)), while when using negative resist (HSQ, MA-N2403, etc.), the solubility decreases (Fig. 11(B)). Finally, the pattern is defined. In the actual process, positive resist ARP6200 was used referring to the substrate material and machining accuracy.

**Appendix C: Analysis of the trend of axial transmission**

Fig. 12 is simulated intensity profile and distributions of excitation PSF (outer SOL), detection PSF (inner SOL) and total PSF along the axis. Here are two important takeaways: First of all, as shown in Fig.12 (G), the focal depth of outer SOL and inner SOL is 240μm and 42μm respectively, while that of total PSF is 42μm. This indicates that inner SOL has a suppression effect on the depth of focus of the system. Secondly, the calculated total PSF focal depth is 1000um, equal to the present value, which shows that the design result is very accurate.

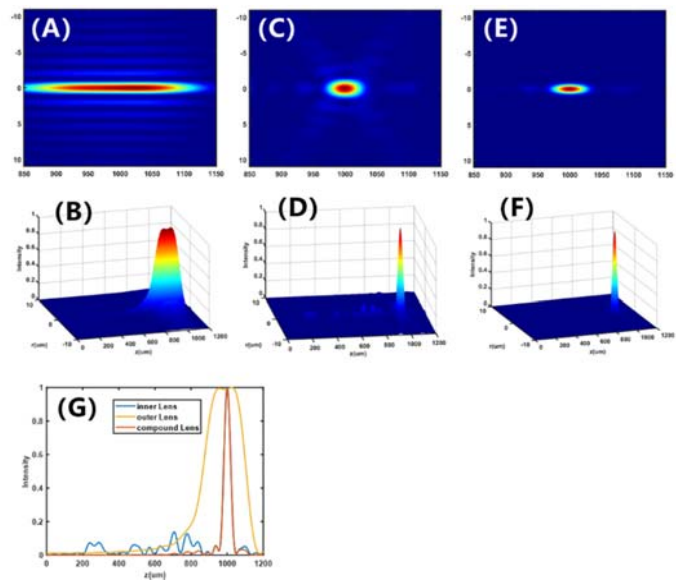

Fig. 12. (A)~(F) Simulate intensity profile of excitation PSF (A, B, outer SOL), detection PSF (C, D, inner SOL) and total PSF (E, F) along the axis. (G) Simulate intensity distributions of excitation PSF (outer SOL), detection PSF (inner SOL) and total PSF along the axis.

**Funding.** National Natural Science Foundation of China (12074444, 61575223); Guangdong Basic and Applied Basic Research Foundation (2021A151501205).

**Disclosures.** The authors declare no conflicts of interest.

**Data availability.** Data underlying the results presented in this paper are not publicly available at this time but may be obtained from the authors upon reasonable request.

**References**

1. S. Hell and J. Wichmann, "Breaking the diffraction resolution limit by stimulated emission: stimulated-emission-depletion fluorescence microscopy," *Opt. Lett.* **19**(11), 780-782(1994).
2. R. Heintzmann and M. Gustafsson, "Subdiffraction resolution in continuous samples," Nat. Photonics **3**(7), 362-364(2009).
3. E. Betzig, G. Patterson, R. Sougrat, O. Lindwasser, S. Olenych, J. Bonifacino, M. Davidson, J. Lippincott-Schwartz, and H. Hess, "Imaging Intracellular Fluorescent Proteins at Nanometer Resolution," Science **313**(5793), 1642-1645(2006).
4. B. Huang, S. A. Jones, B. Brandenburg, and X. Zhuang, "Whole-cell 3D STORM reveals interactions between cellular structures with nanometer-scale resolution," Nat. Methods **5**(12), 1047-1052(2008).
5. Y. Khurgin and Y. Yakovlev, "Progress in the Soviet Union on the theory and applications of bandlimited functions," Proc. IEEE **65**(7), 1005-1029(1977).

6. F. Huang, N. Zheludev, and Y. Chen, "Javier Garcia de Abajo, F. Focusing of Light by a Nano-Hole Array," Appl. Phys. Lett. **90**(9), 091119(2007).

7. K. Huang, H. Ye, J. Teng, S. Yeo, B. Luk'yanchuk, and C. Qiu, "Optimization-free superoscillatory lens using phase and amplitude masks," Laser Photonics Rev. **8**(1), 152-157(2014).

8. Y. Kozawa, D. Matsunaga, and S. Sato, "Superresolution imaging via superoscillation focusing of a radially polarized beam," Optica 5(2), **86**(2018).

9. J. Pendry, "Negative Refraction Makes a Perfect Lens," Phys. Rev. Lett. **85**(18), 3966-3969(2000).

10. N. Yu, P. Genevet, M. Kats, F. Aieta, J. Tetienne, F. Capasso, and Z. Gaburro, "Light propagation with phase discontinuities: Generalized laws of reflection and refraction," Science **334**(6054), 333-337(2011).

11. A. Kildishev, A. Boltasseva, and V. Shalaev, "Planar photonics with metasurfaces," Science **339**(6125), 1232009(2013).

12. Z. Li, M. Kim, C. Wang, Z. Han, S. Shrestha, A. Overvig, M. Lu, A. Stein, A. Agarwal, M. Lončar, and N. Yu, "Controlling propagation and coupling of waveguide modes using phase-gradient metasurfaces," Nat. Nanotechnol. **12**(7), 675-683(2017).

13. S. Divitt, W. Zhu, C. Zhang, H. Lezec, and A. Agrawal, "Ultrafast optical pulse shaping using dielectric metasurfaces," Science **364**(6443), 890-894(2019).

14. X. Ni, A. Kildishev, and V. Shalaev, "Metasurface holograms for visible light," Nat. Commun. **4**(1), 2807(2013).

15. W. Chen, K. Yang, C. Wang, Y. Huang, G. Sun, I. Chiang, C. Liao, W. Hsu, H. Lin, S. Sun, L. Zhou, A. Liu, and D. Tsai, "High-Efficiency Broadband Meta-Hologram with Polarization-Controlled Dual Images," Nano Lett. **14**(1), 225-230(2014).

16. G. Zheng, H. Mühlenbernd, M. Kenney, G. Li, T. Zentgraf, and S. Zhang, "Metasurface holograms reaching 80% efficiency," Nat. Nanotechnol. **10**(4), 308-312(2015).

17. E. Arbabi, S. Kamali, A. Arbabi, and A. Faraon, "Vectorial Holograms with a Dielectric Metasurface: Ultimate Polarization Pattern Generation," ACS Photonics **6**(11), 2712-2718(2019).

18. H. Ren, X. Fang, J. Jang, J. Bürger, J. Rho, and S. Maier, "3D-printed complex-amplitude metasurface for orbital angular momentum holography," Nat. Nanotechnol. **15**, 948-955(2020).

19. F. Aieta, P. Genevet, N. Yu, M. Kats, Z. Gaburro, and F. Capasso, "Out-of-plane reflection and refraction of light by anisotropic optical antenna metasurfaces with phase discontinuities," Nano Lett. **12**(3), 1702-1706(2012).

20. E. Arbabi, A. Arbabi, S. Kamali, Y. Horie, M. Faraji-Dana, and A. Faraon, "MEMS-tunable dielectric metasurface lens," Nat. Commun. **9**(1), 812(2018).

21. Z. Yang, Z. Wang, Y. Wang, X. Feng, M. Zhao, Z. Wan, L. Zhu, J. Liu, Y. Huang, and J. Xia, "Wegener, M. Generalized Hartmann-Shack array of dielectric metalens sub-arrays for polarimetric beam profiling," Nat. Commun. **9**(1), 4607(2018).

22. Q. Fan and Ting. Xu, "Research progress of imaging technologies based on electromagnetic metasurfaces," Acta Phys. Sin. **66**(14), 144208(2017).

23. Z. Zhang, Y. Liang, and T. Xu, "Research advances of hyperbolic metamaterials and metasurfaces," Opto-Electron. Eng. **44**(3), 276-288(2017).

24. H. X. Zhao, J. L. Xie, J. J. Liu, "Optical and acoustic super-resolution imaging in a Stampfli-type photonic quasi-crystal flat lens," Results in Physics 27, 104537(2021).

25. Z. J. Jiang, J. J. Liu, "Progress in far-field focusing and imaging with super-oscillation," Acta Physica Sinica 65(23), 234203 (2016).

26. J. W. Guo, W. Tan, J. L. Xie, J. J. Liu, "One-dimensional photonic quasi-crystal plano-V lens," J. Infrared Millim. Waves 40(5), 589~594(2021).

27. W. Chen, P. Török, M. Foreman, C. Liao, W. Tsai, P. Wu, and D. Tsai, "Integrated Plasmonic Metasurfaces for Spectropolarimetry," Nanotechnology **27**(22), 224002(2016).

28. P. Wu, J. Chen, C. Yin, Y. Lai, T. Chung, C. Liao, B. Chen, K. Lee, C. Chuang, C. Wang, and D. Tsai, "Visible Metasurfaces for On-Chip Polarimetry," ACS Photonics **5**(7), 2568-2573(2017).

29. Z. X. Wu, J. X. Zhu, Y. Y. Zou, H. Deng, L. Xiong, Q. C. Liu, and L. P. Shang, "Superoscillatory metalens for polarization conversion and broadband tight focusing of terahertz waves," Optical Materials,**123**, 111924 (2022).

30. M. Berry and S. Popescu, "Evolution of quantum superoscillations and optical superresolution without evanescent waves," J. Phys. A: Math. Gen. **39**(22), 6965-6977(2006).

31. A. Wong and G. Eleftheriades, "Temporal Pulse Compression Beyond the Fourier Transform Limit," IEEE Trans. Microwave Theory Tech. **59**(9), 2173-2179(2011).

32. A. Wong and G. Eleftheriades, "An Optical Super-Microscope for Far-field, Real-time Imaging Beyond the Diffraction Limit," Sci. Rep. **3**(1), 1715(2013).

33. E. Rogers, S. Savo, J. Lindberg, T. Roy, M. Dennis and N. Zheludev, "Super-oscillatory optical needle," Appl. Phys. Lett. **102**(3), 031108(2013).

34. T. Roy, E. Rogers, G. Yuan, and N. Zheludev, "Point spread function of the optical needle super-oscillatory lens," Appl. Phys. Lett. **104**(23), 231109(2014).

35. X. Dong, A. Wong, and M. Kim, "Eleftheriades, G. Superresolution far-field imaging of complex objects using reduced superoscillating ripples," Optica **4**(9), 1126-1133(2014).

36. V. Le, X. Wang, C. Kuang, and X. Liu, "Background suppression in confocal scanning fluorescence microscopy with superoscillations," Opt. Commun. **426**(1), 541-546(2018).

37. B. Gary, S. Nisan, "Historical development of FINCH from the beginning to single-shot 3D confocal imaging beyond optical resolution," Appl. Opt. **61**(5), B121-B131(2022).

38. E. Rogers, J. Lindberg, T. Roy, S. Savo, J. Chad, M. Dennis, and N. Zheludev, "A super-oscillatory lens optical microscope for subwavelength imaging," Nat. Mater. **11**(5), 432-435(2012).

39. A. Nagarajan, F. Silvestri, G. Gerini, L. P. Stoevelaar, M. Siemons, S. M. B. Bäumer, and V. G. Achanta, "Reflection confocal nanoscopy using a superoscillatory lens," optics express, **27**(14), 20012-20027(2019).

40. G. Chen, Z. Wen, and C. Qiu, "Superoscillation: from physics to optical applications," Light: Sci. Appl. 8(4), 56(2019).

41. E. T. F. Rogers, S. Quraishe, K. S. Rogers, T. A. Newman, P. J. S. Smith, and N. I. Zheludev, "Far-field unlabeled super-resolution imaging with superoscillatory illumination," APL Photonics, **5**(6), 066107(2020).

42. N. Zheludev, "What diffraction limit," Nat. Mater. **7**(6), 420-422(2008).

43. W. Srituravanich, L. Pan, Y. Wang, C. Sun, D. Bogy, and X. Zhang, "Flying plasmonic lens in the near field for high-speed nanolithography," Nat. Nanotechnol. **3**(12), 733-737(2008).

44. F. Huang and N. Zheludev, "Super-resolution without evanescent waves," Nano Lett. **9**(3), 1249-1254(2009).

45. Z. Wen, Y. He, Y. Li, L. Chen, and G. Chen, "Super-oscillation focusing lens based on continuous amplitude and binary phase modulation," Opt. express **22**(18), 22163-22171(2014).

46. E. Rogers and N. Zheludev, "Optical super-oscillations: sub-wavelength light focusing and super-resolution imaging," J. Opt. **15**(9), 4008(2013).

47. L. Novotny and B. Hecht, "Principles of Nano-Optics," Phys. Today **60**(7), 62(2007).

48. X. S. Xie, Y. Chen, K. Yang, and J. Y. Zhou, "Harnessing the Point-Spread Function for High-Resolution Far-Field Optical Microscopy," Phys. Rev. Lett. **113**(26), 263901(2014).

49. X. Xie, K. Yang, and J. Zhou, "Generalized vector wave theory for ultrahigh resolution confocal optical microscopy," J. Opt. Soc. Am. A **34**(1), 61-67(2017).

50. G. Chen, K. Zhang, A. Yu, X. Wang, Z. Zhang, Y. Li, Z. Wen, C. Li, L. Dai, S. Jiang, and F. Lin, "Far-field sub-diffraction focusing lens based on binary amplitude-phase mask for linearly polarized light," Opt. Express **24**(10), 11002-11008(2016).

51. D. Deng, and Q. Guo, "Analytical vectorial structure of radially polarized light beams," Opt. Lett. **32**(18), 2711-2713(2007).

52. W. Carter, "Electromagnetic Field of a Gaussian Beam with an Elliptical Cross Section," J. Opt. Soc. Am. **62**(10), 1195-1201(1972).

53. V. Magni, G. Cerullo, and S. De Silvestri, "High-accuracy fast Hankel transform for optical beam propagation," J. Opt. Soc. Am. A **9**(11), 2031-2033(1992).

54. L. Knockaert, "Fast Hankel transform by fast sine and cosine transforms: the Mellin connection," IEEE Trans. Signal Process. **48**(6), 1695-1701(2000).

55. D. Zhang, X. Yuan, and N. Ngo, "Shum, P. Fast Hankel transform and its application for studying the propagation of cylindrical electromagnetic fields," Opt. Express **10**(12), 521-525(2010).

56. N. Jin, and Y. Rahmat-Samii, "Advances in Particle Swarm Optimization for Antenna Designs: Real-Number, Binary, Single-Objective and Multiobjective Implementations," IEEE Trans. Antennas Propag. **55**(3), 556-567(2007).

57. F. Qin, K. Huang, J. F. Wu, J. F. Teng, C. W. Qiu, and M. H. Hong, "A Supercritical Lens Optical Label-Free Microscopy: Sub-Diffraction Resolution and Ultra-Long Working Distance," Adv. Mater. **29**(8), 1602721(2017).

58. S.W. Hell, P. E. Hhninen, A. Kuusisto, M. Schrader, Erkki Soini, "Annular aperture two-photon excitation microscopy", Opt. Communications **117**(1-2), 20-24 (1995).

59. G. H. Yuan, K. S. Rogers, E. T.F. Rogers, and N. I. Zheludev, "Far-Field Superoscillatory Metamaterial Superlens," Physical review applied, **11**(6), 064016(2019).

60. W. L. Li, P. He, W. Z. Yuan, and Y. T. Yu, "Efficiency-enhanced and sidelobe-suppressed super-oscillatory lenses for sub-diffraction-limit fluorescent imaging with ultralong working distance," Nanoscale. **12**(13), 1-3(2020).